\documentclass{appolb}
\usepackage{epsfig}
\newcommand{\En}{E_n}
\newcommand{\Fn}{F_n}
\newcommand{\Enu}{E_{\overline{\nu}}}

\newcommand{\Enubar}{\epsilon_{\overline{\nu}}}
\newcommand{\thnu}{\overline{\theta}_{\overline{\nu}}}
\newcommand{\cth}{\cos\overline{\theta}_{\overline{\nu}}}
\newcommand{\Fnu}{F_{\overline{\nu}}}
\newcommand{\tbar}{\overline{\tau}_n}

\newcommand{\Enmax}{E_n^{\rm max}}

\newcommand{\eps}{\epsilon}
\newcommand{\nuebar}{\overline{\nu}_e}

\newcommand{\Fnue}{F_{\nu_e}}
\newcommand{\Fnumu}{F_{\nu_\mu}}

\newcommand{\Epho}{E_\gamma}
\newcommand{\Epi}{E_\pi}

\newcommand{\Fpho}{F_\gamma}
\newcommand{\bwide}{\begin{widetext}}
\newcommand{\ewide}{\end{widetext}}
\newcommand{\beq}[1]{\begin{equation} \label{(#1)}}
\newcommand{\eeq}{\end{equation}}
\newcommand{\ba}[1]{\begin{eqnarray} \label{(#1)}}
\newcommand{\ea}{\end{eqnarray}}

\begin{document}
\title{{\bf Exploring the Universe beyond the Photon Window}%
\thanks{Presented at the XXXIV International Symposium on Multiparticle Dynamics 2004}%
}
\author{Luis A. Anchordoqui
\address{Department of Physics, Northeastern University, Boston MA 02115}
%\and
%the Name(s) of other Author(s)
%\address{and their affiliation}
}
\maketitle
\begin{abstract}
\noindent In this talk I review how to identify cosmic ray 
accelerators that are high energy neutrino emitters. I also delineate the 
prospects for a new multi-particle astronomy: neutrons as directional 
pointers + antineutrinos as inheritors of directionality. 

\end{abstract}
\PACS{96.40.-z, 95.85.Ry, 98.70.Sa}
  
\section{Introduction}

\begin{figure}
\label{gamf} 
\centering 
\leavevmode \epsfysize=5.5cm \epsfbox{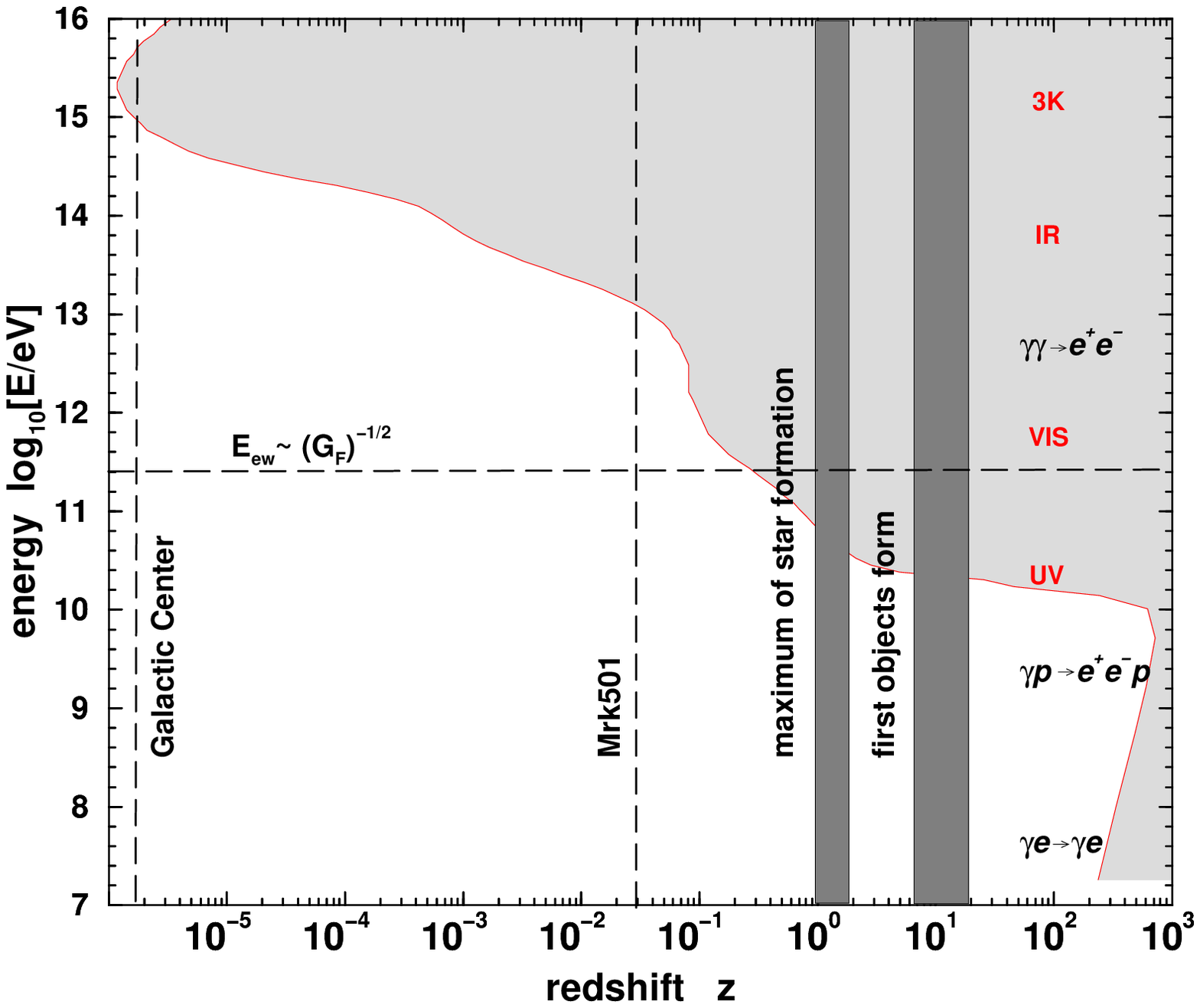}
\leavevmode \epsfysize=5.4cm \epsfbox{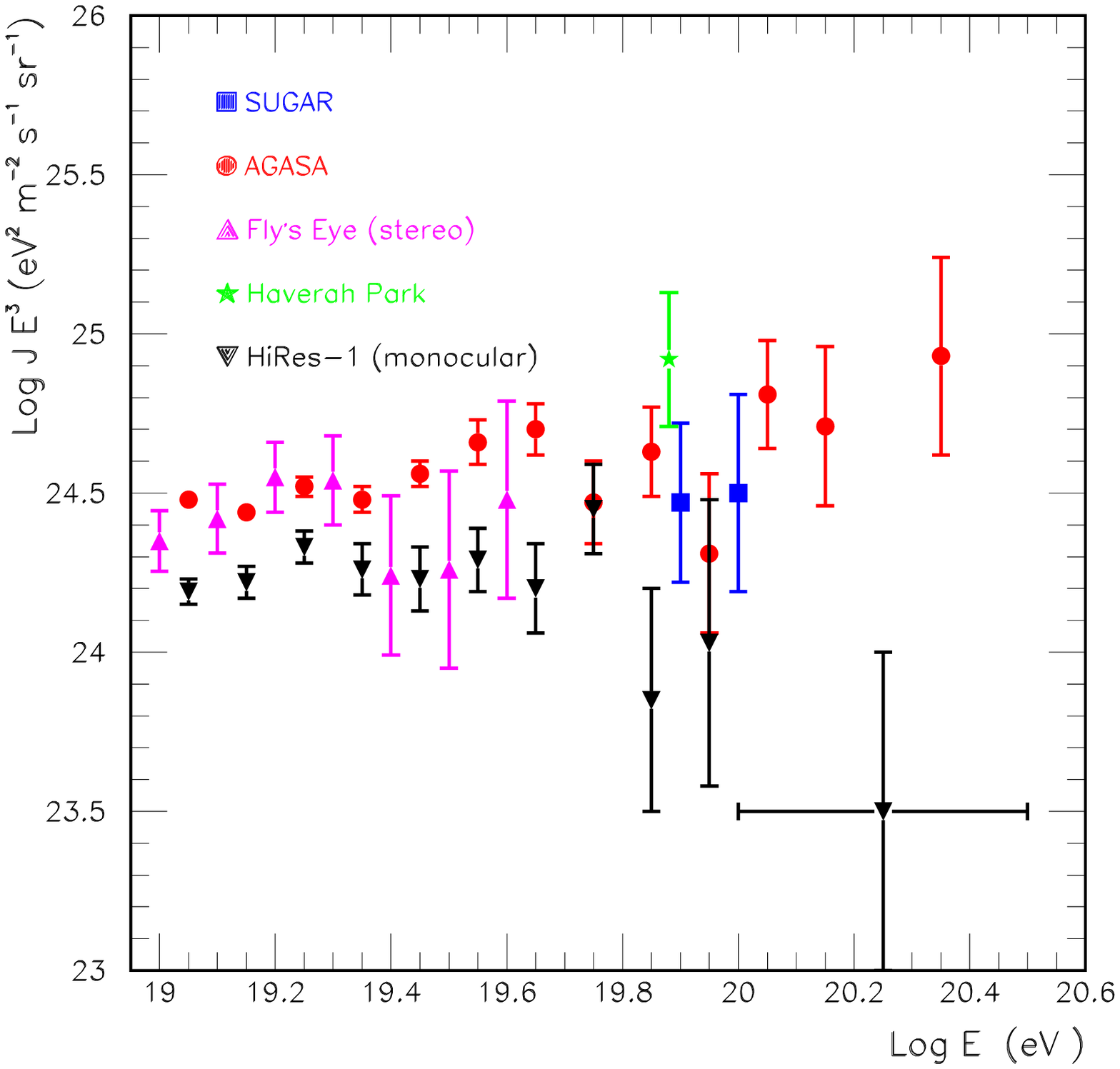}\\ 
\caption{Left: Mean interaction length for photons on the 
ultraviolet, visible, infrared, and microwave backgrounds~\cite{Learned:2000sw}. 
Right:
Upper end of the cosmic ray energy spectrum as observed by 5
different experiments (for details see main text).}
\end{figure}

Conventional astronomy spans about 18 decades in photon
wavelengths, from $10^4$~cm radio-waves to $10^{-14}$~cm
$\gamma$-rays of GeV energy. Because the universe is opaque to photons of
TeV energy and above (see Fig.~\ref{gamf})~\cite{Learned:2000sw}, 
present studies focus on hadrons,
neutrinos, and gravitational waves as messengers probing the high
energy universe. The best candidates to serve as messengers in a
new astronomy of the high energy behavior of distant sources are
neutral particles. This is because the orbit of a charged cosmic
ray can be substantially bent, both by extragalactic magnetic fields~\cite{Sigl:2004gi} 
and by the ambient magnetic field of our own Galaxy~\cite{Stanev:1996qj}; 
destroying the possibility of locating the source.
The most promising messenger is the neutrino: it can be copiously
produced in cosmic beam dumps and can traverse unscathed dense
astrophysical environments. In this talk I delineate the prospects to
identify  high energy neutrino emitters.

\section{Cosmic Ray Astronomy}

For primary energy $> 1$~GeV, the observed cosmic ray intensity can be 
described by a series of power laws with the flux falling about 3 orders 
of magnitude for each decade increase in energy~\cite{Anchordoqui:2002hs}.
In recent years, a somewhat confused picture {\it vis--{\`a}--vis} the 
energy spectrum and arrival direction distribution has been emerging.  
Since 1998, the AGASA Collaboration has consistently 
reported~\cite{Takeda:1998ps} a continuation of the spectrum beyond the 
expected Greisen--Zatsepin--Kuzmin (GZK) cutoff~\cite{Greisen:1966jv}, 
which should arise at about $10^{10.9}$~GeV if cosmic ray sources are
at cosmological distances.  This theoretical feature of the spectrum is mainly a consequence
of interactions of the primary cosmic ray with the microwave background radiation. In contrast, 
the most recent results from HiRes~\cite{Abbasi:2002ta}
describe a spectrum which is consistent with the 
expected GZK feature. The discrepancy between the 2 estimated fluxes 
is shown in Fig.~\ref{gamf} (for comparison, the  intensity as seen by 
the Fly's Eye~\cite{Bird:wp}, the Haverah Park~\cite{Ave:2001hq} and the SUGAR~\cite{Anchordoqui:2003gm} 
experiments is also shown). 
This situation is suggestive of the 
challenge posed by systematic errors in these types of measurements. Further 
confusing the issue, the AGASA Collaboration reports observations 
of event clusters which have a chance probability smaller than 1\% to arise 
from a random distribution~\cite{Hayashida:bc}, whereas 
the recent analysis reported
by the HiRes Collaboration showed that their data are consistent with no
clustering among the highest energy events~\cite{Abbasi:2004ib}.
In my opinion, it is {\it very important} to rigorously define the 
corresponding budget of statistical significance and search criteria 
{\em prior} to studying the data, since defining them {\em a posteriori} 
may inadvertently introduce an indetermimant number of ``trials'' and thus 
make it impossible to assign the correct statistical significance to the 
search result. In this direction,  with the aim of avoiding accidental bias on the
number of trials performed in selecting the angular bin, the original
claim of the AGASA Collaboration was re-examined
considering
only those events observed after the original claim~\cite{Finley:2003ur}. 
This study
showed that the evidence for clustering in the AGASA data set is weaker than was
previously supposed, and is consistent with the hypothesis of isotropically
distributed arrival directions. The confusing experimental situation regarding the GZK 
feature should be resolved in the near future by 
the Pierre Auger Observatory~\cite{Abraham:2004dt}, which will provide not only a data set of 
unprecedented size, but also the machinery for controlling some of the more problematic systematic 
uncertainties.   

With the controversy over the GZK cutoff, one may have missed the fact 
that below $10^{10}$~GeV AGASA has revealed a correlation of the arrival 
direction of the cosmic
rays to the Galactic Plane (GP) at the $4\sigma$
level~\cite{Hayashida:1998qb}. The GP excess, which is roughly 4\% of
the diffuse flux, is mostly
concentrated in the direction of the Cygnus region, with a second
spot towards the Galactic Center (GC). The anisotropy signal spans a narrow energy window, from 
$10^{8.7}$~GeV to $10^{9.3}$~GeV. Evidence at
the 3.2$\sigma$ level for GP
enhancement in a similar energy range has also been reported by the Fly's Eye
Collaboration~\cite{Bird:1998nu}. Interestingly, the full Fly's Eye data include a directional signal from 
the Cygnus region which was somewhat lost in an unsuccessful attempt to relate it to $\gamma$--ray emission 
from Cygnus X--3~\cite{Cassiday:kw}.\footnote{The complete isotropy at PeV energies revealed by KASCADE 
data~\cite{Antoni:2003jm} vitiate direction-preserving photons as primaries.} Additionally, 
a 3.4$\sigma$ excess from an 
extended region surrounding Cygnus X-3 has been reported by the Yakutsk Collaboration~\cite{Glushkov}. 
On the other hand, the existence of a point-like excess in the
direction of the GC has been confirmed via independent
analysis~\cite{Bellido:2000tr} of data collected with the SUGAR facility. This is a remarkable level of
agreement among experiment using a variety of techniques.

Independent evidence may be emerging for a cosmic
accelerator in the Cygnus spiral arm. The HEGRA experiment has detected an
extended TeV $\gamma$-ray source in the Cygnus region with no clear
counterpart and a spectrum, 
\begin{equation}
\frac{d\Fpho}{d\Epho} = 4.7^{\pm 2.1}_{\pm 1.3}
\times 10^{-13}
\left(\frac{E_\gamma}{1~{\rm TeV}}\right)^{-1.9^{\pm 0.3}_{\pm 0.3}} 
\,\,{\rm cm}^{-2} {\rm s}^{-1} {\rm TeV}^{-1} \,\,,
\label{1}
\end{equation}
not easily accommodated with synchrotron radiation
by electrons~\cite{Aharonian:2002ij}.  Especially intriguing is the 
possible association of this
source with Cygnus-OB2, a cluster of more than 2700 (identified) young,
hot stars with a total mass of $\sim 10^4$ solar
masses~\cite{Knodlseder:2000vq}. Proton
acceleration to explain the TeV photon signal requires only 0.1\% efficiency
for the conversion of the energy in the stellar wind into cosmic ray
acceleration. Also, the stars in Cygnus-OB2 could be the origin of
time-correlated, clustered supernova remnants forming a source of cosmic
ray nuclei.  {\sl An
immediate consequence of this picture is the
creation of free neutrons via nuclei photodisintegration on
background photon fields.} These liberated neutrons are presumably
responsible for the observed directional signals. {\sl This
implies that it may not be a coincidence that the signal appears
first at energies where the neutron lifetime allows propagation
distances of galactic scales, i.e., 10 kpc.}

\section{Neutrino Astronomy}

Neutrinos can serve as unique
astronomical messengers. Except for oscillations induced by
transit in a vacuum higgs field, neutrinos propagate without
interactions between source and Earth, providing powerful probes
of high energy astrophysics~\cite{Gaisser:1994yf}. 
The deployment of under-ice/water telescopes in the Northern~\cite{Montaruli} and 
Southern~\cite{Ahrens:2003ix} hemispheres will greatly increase the statistics required 
for the realization of such a program. In this section 
we study possible mechanisms for neutrino production in astrophysical sources 
and discuss how flavor oscillations can distort the initial injection spectra.

\subsection{Flavor metamorphosis}

In recent years, stronger and stronger experimental evidence for neutrino oscillations has been 
accumulating. Neutrino flavor change implies: {\it (i)} that neutrinos have nonzero masses, i.e., 
there is a spectrum of 3 or more neutrino mass eigenstates that are analogues of the charged lepton 
mass eigenstate $l_\alpha$ $(\alpha = e, \mu, \tau)$; {\it (ii)} leptonic mixing, i.e., the weak 
interaction coupling the $W$ boson to a charged lepton and a neutrino, can also couple any charged lepton 
mass eigenstate $l_\alpha$ to any neutrino mass eigenstate $\nu_j$ ($j = 1, 2, 3, \dots$).  The 
superposition of neutrino mass 
eigenstates produced in association with the charged lepton of flavor $\alpha,$
\begin{equation}
|\nu_\alpha\rangle = \sum_j U_{\alpha j}^* |\nu_j\rangle \,\,,
\label{state}
\end{equation}
is the state we refer to as the neutrino 
of flavor alpha, where $U_{\alpha j}^*$ are elements of the neutrino mass-to-flavor mixing matrix, 
fundamental to particle physics.  

Throughout this talk we consider 3 neutrino species. In this case,   
atmospheric data~\cite{Fukuda:1998mi} indicate that $\nu_\mu$ and $\nu_\tau$ are 
maximally mixed and reactor data~\cite{Bilenky:2001jq} points to $|U_{e3}|^2 \ll 1$. Thus, to simplify 
the discussion  hereafter we use the fact that $|U_{e3}|^2$ is nearly zero 
to ignore CP violation and assume real matrix elements. With this in mind, 
one can define a mass basis as follows,
\begin{equation}
|\nu_1 \rangle = \sin \theta_\odot |\nu^\star\rangle +  \cos \theta_\odot |\nu_e\rangle \,\,,
\end{equation}
\begin{equation}
|\nu_2 \rangle =  \cos \theta_\odot |\nu^\star\rangle  -\sin \theta_\odot |\nu_e\rangle \,\,,
\end{equation}
and
\begin{equation}
|\nu_3 \rangle = \frac{1}{\sqrt{2}} (|\nu_\mu \rangle + |\nu_\tau \rangle) \,\,,
\label{3rd}
\end{equation}
where $\theta_\odot$ is the solar mixing angle and 
\begin{equation}
|\nu^\star\rangle = \frac{1}{\sqrt{2}} (|\nu_\mu\rangle - |\nu_\tau \rangle)
\label{orthogonal}
\end{equation}
is the eigenstate orthogonal to $|\nu_3 \rangle.$ Inversion of the neutrino mass-to-flavor 
mixing matrix leads leads to
\begin{equation}
|\nu_e \rangle = \cos \theta_\odot |\nu_1\rangle - \sin \theta_\odot |\nu_2 \rangle
\end{equation}
and
\begin{equation}
|\nu^\star \rangle = \sin \theta_\odot |\nu_1\rangle + \cos \theta_\odot |\nu_2 \rangle \,\,.
\end{equation}
Finally, by adding Eqs.~(\ref{3rd}) and (\ref{orthogonal}) one obtains the $\nu_\mu$ flavor eigenstate,
\begin{equation}
|\nu_\mu \rangle = \frac{1}{\sqrt{2}} \left[ |\nu_3 \rangle + \sin \theta_\odot |\nu_1 \rangle + 
\cos \theta_\odot |\nu_2\rangle \right] \,\,,
\end{equation}
and by substracting these same equations the $\nu_\tau$ eigenstate.

The evolution in time of the $\nu_i$ component of a neutrino initially born as $\nu_\alpha$ in the rest 
frame of that 
component is described by  Schr\"odinger's equation,
\begin{equation}
|\nu_i (\tau_i) \rangle = e^{-i m_i \tau_i} |\nu_i (0)\rangle \,\,,
\end{equation}
where $m_i$ is the mass of $\nu_i$ and $\tau_i$ is the proper time. 
In the lab frame, the Lorentz invariant phase factor may be written as $e^{-i(E_i t - p_i L)}$, where $t,$ 
$L, $ $E_i,$ and $p_i,$ are respectively, the time, the position, the energy, and the momentum of $\nu_i$ 
in the lab frame. Since the neutrino is extremely 
relativistic $t \approx L$ and $E_i = \sqrt{p^2 + m_i^2} \approx p + m_i^2/2p$.  Hence, from 
Eq.~(\ref{state}) it follows that the state vector of a neutrino born as $\nu_\alpha$ after propagation 
of distance $L$ becomes
\begin{equation}
|\nu_\alpha(L) \rangle \approx \sum_i U_{\alpha i} \, e^{-i (m_i^2/2E)\,\, L}\,\,|\nu_i \rangle \,\,,
\label{ritazza}
\end{equation}
where $E \approx p$ is the average energy of the various mass eigenstate components of the neutrino.
Using the unitarity of $U$ to invert Eq.~(\ref{state}), from Eq.~(\ref{ritazza}) one finds that 
\begin{equation}
|\nu_\alpha(L) \rangle \approx \sum_\beta \left[\sum_i U_{\alpha i} \, e^{-i (m_i^2/2E)\,\, L}\,\,U_{\beta i}\,\,\right]\,|\nu_\beta \rangle \,\,.
\end{equation}
In other words, 
the propagating mass eigenstates acquire relative 
phases giving rise to flavor oscillations. Thus, 
after traveling a distance $L$ an initial state 
$\nu_\alpha$ becomes a superposition of all flavors, with probability of transition to flavor $\beta,$ 
$|\langle \nu_\beta| \nu_\alpha(L) \rangle|^2,$ given by
\begin{equation}
P(\nu_\alpha \to \nu_\beta) = \delta_{\alpha \beta} - 4 \sum_{i>j} U_{\alpha i}\, U_{\beta i}\, 
U_{\alpha j} \, U_{\beta j} \, \sin^2 \Delta_{ij}\,\,,
\end{equation}
where $\Delta_{ij} \sim \delta m_{ij}^2 L/ 2E,$ and $\delta m_{ij}^2 = m_i - m_j.$

For $\Delta_{ij} \gg 1$, the phases will be erased by uncertainties in $L$ and $E$. Consequently,
averaging over $\sin^2 \Delta_{ij}$ one finds
\begin{equation}
P(\nu_\alpha \to \nu_\beta) = \delta_{\alpha \beta} - 2 \sum_{i>j} U_{\alpha i}\, U_{\beta i}\, 
U_{\alpha j} \, U_{\beta j} \,.
\label{paco}
\end{equation}
Now, using $2 \sum_{1>j} = \sum_{i,j} - \sum_{i=j},$ Eq.~(\ref{paco}) can be re-written as
\begin{eqnarray}
P(\nu_\alpha \to \nu_\beta) & = & \delta_{\alpha \beta} -  \sum_{i,j} U_{\alpha i}\, U_{\beta i}\, 
U_{\alpha j} \, U_{\beta j} \, +  \sum_{i} U_{\alpha i}\, U_{\beta i}\, 
U_{\alpha i} \, U_{\beta i} \nonumber \\
 & = & \delta_{\alpha \beta} - \left( \sum_{i} U_{\alpha i}  U_{\beta i} \right)^2 + \sum_{i}   
U_{\alpha i}^2  U_{\beta i}^2\,.
\label{PP}
\end{eqnarray}
Since $\delta_{\alpha \beta}$ = $\delta_{\alpha \beta}^2,$ the first and second terms in Eq.~(\ref{PP}) 
cancel each other, yielding
\begin{equation}
P(\nu_\alpha \to \nu_\beta) = \sum_{i} U_{\alpha i}^2 \,\,U_{\beta i}^2 \,\,. 
\end{equation}
The probabilities for flavor oscillation are then
\begin{equation}
P(\nu_\mu \to \nu_\mu) = P(\nu_\mu \to \nu_\tau)= \frac{1}{4}\, (\cos^4 \theta_\odot + \sin^4 \theta_\odot + 1) \,\,,
\label{p1}
\end{equation}
\begin{equation}
P(\nu_\mu \to \nu_e) = P(\nu_e \to \nu_\mu) = P(\nu_e \to \nu_\tau) = \sin^2 \theta_\odot \,\, \cos^2 \theta_\odot \,\,,
\label{p2}
\end{equation} 
and
\begin{equation}
P(\nu_e \to \nu_e) = \cos^4 \theta_\odot + \sin^4 \theta_\odot \,\,.
\label{p3}
\end{equation}

Now, let the ratios of 
neutrino flavors at production in the cosmic sources 
be written as $w_e : w_\mu : w_\tau$ with $\sum_\alpha w_\alpha = 1,$ so that the 
relative fluxes of each  mass eigenstate  are given by 
$w_j = \sum_\alpha \omega_\alpha \,\,U_{\alpha j}^2$. From our previous discussion, we conclude that 
the probability of measuring on Earth a flavor $\alpha$ is given by
\begin{equation}
P_{\nu_\alpha \,\,{\rm detected}} = \sum_j U_{\alpha j}^2 \,\, \sum_\beta w_\beta \,\,U_{\beta j}^2 \,\,.
\end{equation}
Straightforward calculation shows that any initial flavor ratio that contains $w_e = 1/3$ will arrive 
at Earth with equipartition on the three flavors.

\subsection{The $\gamma$--$\nu$ connection}

There are two principal mechanisms for TeV gamma ray production:
{\it (i)}~Electrons  undergo bremsstrahlung in the
magnetic field and/or inverse Compton scattering in the ambient
photon sea or {\it (ii)}~the gamma rays are directly traced to $\pi^0$
decay.\footnote{Pions can be produced in $pp$ and/or $p\gamma$ collisions. In this talk we 
focus attention on the former; a generalization to the photopion production process is 
straightforward, see~\cite{Anchordoqui:2004eb} for details.}  
Only the second scenario can accommodate baryonic cosmic ray
production. Since such cosmic rays are observed, it is reasonable to 
assume that at least some  gamma ray sources operate according to the 
second mechanism.

Inelastic $pp$ collisions lead to roughly equal numbers of $\pi^0$'s, 
$\pi^+$'s, and $\pi^-$'s, hence one expects two photons, two $\nu_e$'s, 
and four $\nu_\mu$'s per $\pi^0$.  On average, the photons carry one-half of
the energy of the pion. 
The average  neutrino energy from the direct pion decay  is
$\langle E_{\nu_\mu} \rangle^\pi = (1-r)\,E_\pi/2 \simeq 0.22\,E_\pi$ 
and that of the muon is $\langle E_{\mu} \rangle^\pi = (1+r)\,E_\pi/2 \simeq 
0.78\,E_\pi$, where $r$ is the ratio 
of muon to the pion mass squared. Now, taking the $\nu_\mu$ from muon decay 
to have 1/3 the energy of the muon, the average energy of the $\nu_\mu$ from 
muon decay is $\langle E_{\nu_\mu} \rangle^\mu =(1+r)E_\pi/6=0.26 \, E_\pi$. 
This gives a total $\nu_\mu$ energy per charged pion
 $ \langle E_{\nu_\mu} \rangle \simeq 0.48 \, E_\pi$, 
with a total $\langle E_{\nu_\mu} \rangle^{\rm total} = 0.96 \langle 
E_\gamma \rangle$ for each triplet of
$\pi^+,$ $\pi^-,$ and  $\pi^0$ produced. For simplicity, hereafter we 
consider that all neutrinos carry one-quarter of the energy of the pion.

The total number of $\gamma$-rays in the energy interval $(E_1/2,\, E_2/2)$ is equal 
to the total number of charged pions in the interval $(E_1,\, E_2)$ and twice the number 
of neutral pions in the same energy interval, 
\begin{equation}
\int_{E_1/2}^{E_2/2} \frac{dF_\gamma}{dE_{\gamma}} dE_\gamma = 2
\int_{E_1}^{E_2} \frac{dF_{\pi^0}}{dE_{\pi}} dE_\pi = 
2 N_{\pi^0}\,\, .
\label{po}
\end{equation}
Additionally, since $N_{\pi^\pm} = 2 \,N_{\pi^0}$, the number of $\nu_\mu$ in 
the energy interval  $(E_1/4,\, E_2/4)$ scales as
\begin{equation}
\int_{E_1/4}^{E_2/4} \frac{dF_{\nu_\mu}}{dE_{\nu}} dE_\nu = 2 
\int_{E_1}^{E_2} \frac{dF_{\pi^\pm}}{dE_{\pi}} dE_\pi =  
2 N_{\pi^\pm}\,.
\label{pu}
\end{equation}
Now, taking $d/dE_2$ on each side of Eqs.~(\ref{po}) and (\ref{pu}) leads to
\begin{equation}
\left. \frac{1}{2} \frac{dF_\gamma}{dE_\gamma} \right|_{E_\gamma = E_2/2} = \left. 2\,
\frac{dF_\pi^0}{dE_\pi} \right|_{E_2} \,\,\,{\rm and} \,\,\,
\left. \frac{1}{4} \frac{dF_{\nu_\mu}}{dE_{\nu}} \right|_{E_\nu = E_2/4} 
= 2 \left. \frac{dF_\pi^\pm}{dE_\pi} \right|_{E_2} \,, 
\end{equation}
respectively. The
energy-bins $dE$ scale with these fractions, and we arrive at
\begin{eqnarray}
\left. \frac{d\Fpho}{d\Epho} \right|_{\Epho=\Epi/2} & = &
  \left.  4\,\frac{dF_{\pi}}{d\Epi} \right|_{\Epi}\,, \nonumber \\
\left. \frac{d\Fnue}{d E_\nu} \right|_{E_\nu = \Epi/4}  & = &
  \left.  8\,\frac{dF_{\pi}}{d\Epi} \right|_{\Epi}\,, \\
\left. \frac{d\Fnumu}{dE_\nu} \right|_{E_\nu= \Epi/4} & = &
\left.    16\,\frac{dF_{\pi}}{d\Epi}\right|_{\Epi}\,,\nonumber
\end{eqnarray}
for the total fluxes at the source, where $\pi$ denotes any one of the
three pion charge-states. 
In propagation to Earth a distance longer than all oscillation lengths, 
flavor changing amplitudes are replaced by probabilities. Using Eqs.~(\ref{p1}), 
(\ref{p2}), and (\ref{p3}) one can check that the 
initial flavor ratio $1:2:0$ mutates in a nearly identical flux for each of 
the three neutrino flavors which is equal to~\cite{Anchordoqui:2004eu}
\begin{equation}
\left. \frac{dF_{\nu_\alpha}}{dE_\nu} \right|_{E_\nu= \Epho/2} = 
\left. 2\frac{d\Fpho}{d\Epho}\right|_{\Epho}\,.
\label{fnu}
\end{equation}

IceCube is, perhaps, the most promising route for neutrino
detection~\cite{Ahrens:2003ix}. This telescope will consist of 80
kilometer-length strings, each instrumented with 60 10-inch
photo-multipliers spaced by 1.7 m. The deepest module is 2.4 km
below the ice surface. The strings are arranged at the apexes of
equilateral triangles 125 m on a side. The instrumented detector
volume is a cubic kilometer. A surface air
shower detector, IceTop, consisting of 160 Auger-style
\v{C}erenkov detectors deployed over $A_{\rm eff} \approx 1$~km$^2$ 
above IceCube, augments the deep-ice component by providing a tool for
calibration, background rejection and air-shower physics. Muons
can be observed from $10^{2}$~GeV to $10^{9}$~GeV. Cascades,
generated by $\nu_e,$ $\overline\nu_e$, $\nu_\tau,$ and $\overline
\nu_\tau$ can be observed above $10^{2}$~GeV and reconstructed at
energies somewhat above $10^{4}$~GeV. For $\nu_\mu$'s of TeV energy, 
the angular resolution $\approx 0.7^\circ$ allows a point source search window of
$\Omega_{1^\circ \times 1^\circ} \approx 3 \times 10^{-4}$~sr.

The Crab Nebula is generally taken as the standard candle of steady TeV 
$\gamma$-ray emission. In the energy $1~{\rm TeV} 
< E_\gamma < 20~{\rm TeV},$ the Crab data can be fitted by a single power 
law~\cite{Aharonian:2000pz},
\begin{equation}
\frac{d\Fpho}{d\Epho} = 2.79^{\pm 0.02}_{\pm 0.5} \times 10^{-7} \,\,
\left(\frac{E_\gamma}{{\rm TeV}}\right)^{-2.59^{\pm 0.03}_{\pm 0.05}} \,\, 
{\rm m}^{-2} \,{\rm 
s}^{-1} \, {\rm TeV}^{-1}\,\,.
\end{equation}
 Even though  the energy spectrum of the Crab Nebula as measured by the HEGRA system 
is in very good agreement with calculations of inverse Compton scattering~\cite{Atoyan}, 
it is interesting and desirable to have an independent observational discriminator 
between these 
two scenarios. For $E_{\nu}^{\rm min} \simeq 1$~TeV, the number 
of $\nu_\mu +\bar \nu_\mu$ showers expected to be detected at IceCube is given by
\begin{equation}
\label{signal}
\left. \frac{dN}{dt}\right|_{\rm signal}  =  A_{\rm eff}
\,\int_{E_{\nu}^{\rm min}}
dE_\nu\,\frac{dF_{\nu_\mu}}{dE_\nu}(E_\nu)\,\,p(E_\nu)\, ,
\end{equation}
where 
$p (E_\nu) \approx 1.3 \times 10^{-6} \,(E_\nu/{\rm TeV})^{0.8}$
denotes the probability (generic to ice/water detectors)
that a $\nu$ (or $\bar \nu$) with energy $E_\nu$ on a
trajectory through the detector produces a signal~\cite{Gaisser:1994yf}.
Using Eq.~(\ref{fnu}) straightforward calculation shows that 
$dN/dt|_{\rm signal} \sim$ 11 yr$^{-1}$~\cite{Guetta:2002hv}. The
event rate of the atmospheric $\nu$-background is
\begin{equation}
\left. \frac{dN}{dt}\right|_{{\rm background}} = A_{\rm eff}\,
\int_{E_{\nu}^{\rm min}} dE_\nu
\,J_{\nu + \bar \nu}(E_\nu)\,\, p (E_\nu)
\,\,\Delta \Omega_{1^\circ \times 1^\circ} \approx 1.5~{\rm yr}^{-1}\,, \label{background}
\end{equation}
where $J_{\nu + \bar \nu} (E_\nu)$ is the
$\nu_\mu + \bar \nu_\mu$ atmospheric flux in the direction of the
Crab region (about $22^\circ$
below the horizon)~\cite{Lipari:hd}. We conclude that the neutrino signal can be easily isolated from 
background at IceCube. Therefore, the hadronic nature of the high energy 
emission from the Crab Nebula can be confirmed or disproved in a few years of operation.

\subsection{The $n$--$\overline\nu$ connection}

Cosmic ray experiments have identified an
excess from the region of the
GP in a limited energy range around $10^{9}$~GeV.
This is very suggestive of neutrons as candidate primaries, because the directional signal 
requires relatively-stable neutral primaries, and time-dilated neutrons can reach the Earth from typical 
Galactic distances when the neutron energy exceeds $10^{9}$~GeV.
In what follows we show that if the Galactic messengers are neutrons,
then those with energies below an $10^9$~GeV
will decay in flight, providing a flux of cosmic
antineutrinos above a TeV which would be {\it observable}
at IceCube.

The basic formula that relates the neutron flux at the 
source ($d\Fn/d\En$) to the antineutrino flux observed at 
Earth ($d\Fnu/d\Enu$) is~\cite{Anchordoqui:2003vc}:
\begin{eqnarray}
\frac{d\Fnu}{d\Enu}(\Enu) & = &
\int d\En\,\frac{d\Fn}{d\En}(\En)
\left(1-e^{-\frac{D\,m_n}{\En\,\tbar}}\right)\,
\int_0^Q d\Enubar\,\frac{dP}{d\Enubar}(\Enubar) \nonumber \\
 & \times & \int_{-1}^1 \frac{d\cth}{2}
\;\delta\left[\Enu-\En\,\Enubar\,(1+\cth)/m_n\right]
\,.
\label{nuflux}
\end{eqnarray}
The variables appearing in Eq.~(\ref{nuflux}) are the antineutrino and
neutron energies in the lab ($\Enu$ and $\En$),
the antineutrino angle with respect to the direction of the
neutron  momentum, in the neutron rest-frame ($\thnu$),
and the antineutrino energy in the neutron rest-frame
($\Enubar$).  The last three variables are not observed
by a laboratory neutrino-detector, and so are integrated over.
The observable $\Enu$ is held fixed.
The delta-function relates the neutrino energy in the lab to the
three integration variables. The parameters appearing in Eq.~(\ref{nuflux}) 
are the
neutron mass and rest-frame lifetime ($m_n$ and $\tbar$). Finally, 
$dP/d\Enubar$ is the
normalized probability that the
decaying neutron produces a $\overline \nu$ with
energy $\Enubar$  in the neutron rest-frame. Note that the maximum $\overline \nu$ energy 
in the neutron 
rest frame is very nearly  $Q \equiv m_n - m_p - m_e = 0.71$~MeV.
Integration of Eq.~(\ref{nuflux})  can be 
easily accomplished, especially when two good approximations are 
applied.
The first approximation is to think of the $\beta$--decay as a 
$1 \to 2 $ process of 
$\delta m_N \to e^- + \overline \nu,$ in which the neutrino is produced 
monoenergetically in the rest frame, with $\epsilon_{\overline \nu} = \epsilon_0 
\simeq \delta m_N (1  - m_e^2/ \delta^2 m_N)/2 \simeq 0.55$~MeV, where 
$\delta m_N \simeq 1.30$~MeV 
is the neutron-proton mass difference. In the lab,
the ratio of the maximum $\overline \nu$ energy to the neutron energy  
is $2 \epsilon_0/m_n \sim 10^{-3},$ 
and so the boosted $\overline \nu$ has a spectrum with 
$E_{\overline\nu} \in (0, 10^{-3} \, E_n).$ 
 The second approximation is to replace the neutron decay probability 
$1 - e^{-Dm_n/E_n \overline \tau_n}$
with a step function $\Theta (E_n^{\rm max} - E_n)$ at some energy 
$E_n^{\rm max} \sim {\cal O}(D \, m_n/\overline{\tau}_n) = 
(D/10~{\rm kpc}) \times 10^{9}$~GeV. 
Combining these two approximations we obtain
\begin{equation}
\label{nuflux4}
\frac{d\Fnu}{d\Enu}(\Enu)=\frac{m_n}{2\,\eps_0}
\int^{\Enmax}_{\frac{m_n\,\Enu}{2\,\eps_0}} \frac{d\En}{\En}\,
    \frac{d\Fn}{d\En}(\En)\,.
\end{equation}
Normalization to the observed ``neutron'' excess at $\sim
10^{9}$~GeV leads via Eq.~(\ref{signal}) to about 20 antineutrino showers per
year~\cite{Anchordoqui:2003vc}.

A direct $\nuebar$ event in IceCube
will make a showering event with poor angular resolution.
Fortunately, neutrino oscillations rescue the signal.
Since the distance to Cygnus OB2 ($D \approx 1.7$~kpc) greatly exceeds the
$\nuebar$ oscillation length
$\lambda_{\rm osc} \sim (E_{\overline \nu}/{\rm PeV}) \times 10^{-2}$~parsecs
(taking the solar oscillation scale
$\delta m^2 \sim 10^{-5}{\rm eV}^2$),
the antineutrinos decohere in transit. Replacing into Eq.~(\ref{p2}) 
the most recent SNO result for the solar mixing angle
$\theta_\odot \simeq 32.5^\circ$~\cite{Ahmed:2003kj}, it is easily seen
that the  arriving antineutrinos are distributed over flavors,
with the muon antineutrino flux $F_{\bar \nu_\mu}$
given by the factor
$\sin^2 (2\,\theta_\odot)/4 \simeq 0.20$
times the original $F_{\bar \nu_e}$ flux.
The $\bar \nu_\tau$ flux is the same,
and via Eq.~(\ref{p3}) the $\bar \nu_e$ flux is 0.6 times the original flux.
The Cygnus region is about $40^\circ$ below the horizon, hence straightforward calculation 
yields a yearly atmospheric background of 1.5 events. Finally, on the basis that HEGRA observations 
given in Eq.~(\ref{1}) emerge from pion decay, one can first estimate the 
accompanying neutrino flux using Eq.~(\ref{fnu}), and then through Eq.~(\ref{signal}) verify that the event rate associated 
with the unidentified HEGRA source is even smaller than the atmospheric background.

In summary, in a few years of observation, IceCube will attain  $5\sigma$
sensitivity for discovery of the Fe$ \rightarrow n \rightarrow
\nuebar\rightarrow \overline \nu_\mu$ cosmic beam, providing the
``smoking ice'' for the GP neutron hypothesis.

\section{Concluding Remarks}

We have analyzed the possibility of detecting the neutrino counterparts of TeV $\gamma$-ray 
observations in a model where $\gamma$-rays originate through $\pi^0$-decay at the source. 
We have found that IceCube will attain sensitivity to observed neutrinos from the Crab Nebula. 
The discussion presented here can be easily generalized to other sources 
(see {\it e.g.,}~\cite{Anchordoqui:2004eu,Alvarez-Muniz:2002tn}). Indeed 
the AMANDA-II experiment has already achieved the 
sensitivity to probe the blazar Mrk 501 in its 1997 flaring state if the neutrino 
and $\gamma$-ray fluxes are equal~\cite{Ahrens:2003pv}. We have also estimated the 
``essentially guaranteed'' $\overline \nu$ 
flux  originated in the decay of neutrons emitted by the Cygnus OB association. The expected 
event rate above 1 TeV at IceCube is $dN/dt|_{\rm signal} \sim 20$~yr$^{-1}$ antineutrino showers 
(all flavors) and a $1^\circ$ directional signal of 4 $\overline \nu_\mu$ events, well above atmospheric 
background.

The detection of (anti)neutrinos pointing towards TeV $\gamma$-ray sources or to the 
$10^{9}$~GeV intriguing directional signals would not only provide a final answer to the origin of these 
particles, but also confirm the acceleration of protons/nuclei to ultrahigh energies.

\section*{Acknowledgments}

This talk is based on several works done in collaboration with Haim Goldberg, 
Francis Halzen, and Tom Weiler. Special thanks go to John Learned for permission to 
reproduce Fig.~\ref{gamf}. I'm also indebted to Bill Gary and Lynn Cominsky for organizing 
a very fruitful conference. This work has been partially supported by the US 
National Science Foundation (NSF) under grant No. PHY-0140407.

\end{document}